# Interpretive Themes in Quantum Physics: Curriculum Development and Outcomes

Charles Baily and Noah D. Finkelstein

*Department of Physics, University of Colorado, Boulder, CO 80309, USA*

**Abstract.** A common learning goal for modern physics instructors is for students to recognize a difference between the experimental uncertainty of classical physics and the fundamental uncertainty of quantum mechanics. Our prior work has shown that student perspectives on the physical interpretation of quantum mechanics can be characterized, and are differentially influenced by the myriad ways instructors approach interpretive themes in their introductory courses. We report how a transformed modern physics curriculum (recently implemented at the University of Colorado) has positively impacted student perspectives on quantum physics, by making questions of classical and quantum reality a central theme of the course, but also by making the beliefs of students (and not just those of scientists) an explicit topic of discussion.

**Keywords:** Physics education research, quantum mechanics, modern physics, curriculum development
**PACS:** 01.40.Fk, 03.65.Aa, 01.40.Di

## INTRODUCTION

Like expert physicists, introductory students differ in their physical interpretations of quantum mechanics. [1–4] We have previously shown how the intuitively *realist* (classical) perspectives of modern physics students can significantly influence their stances on questions central to the so-called *measurement problem*: Is the wave function physically real, or simply a mathematical tool? Does the *collapse of the wave function* represent a change in information, or a physical transition not described by any equation? Do electrons exist as localized particles at all times? [2] These questions are of both personal and academic interest to students, but are mostly only superficially addressed in introductory courses, often for fear of generating further confusion in an already abstract and challenging topic area. We have found that introductory students are indeed capable of developing sophisticated and nuanced stances on such interpretive questions, but are often lacking the conceptual resources to articulate their beliefs. [2, 4]

Our prior work has sought to understand and characterize student perspectives on the physical interpretation of quantum mechanics, and in doing so, we have demonstrated various impacts on student thinking from myriad instructional approaches with respect to interpretive themes. [3] Interpretive themes in quantum mechanics are an often *hidden* aspect of modern physics instruction, according to three criteria: A) Students develop stances on these interpretive themes, regardless of whether instructors adequately attend to them; B) Those beliefs tend to be more novice-like (intuitively *realist*) in contexts where instruction is less explicit; and C) Explicit instruction is typically not meaningful for students beyond the specific contexts in which they arise. [3, 4]

We have recently implemented further research-based transformations to an introductory modern physics curriculum developed at the University of Colorado, with an aim to have students not only be consciously aware of their own (often intuitive and tacit) beliefs about classical and quantum reality, but also for them to acquire the necessary language and tools to identify and articulate those beliefs in a variety of contexts. We describe in this paper the nature of this transformed curriculum, and show how it has positively impacted students' personal interest in quantum mechanics, and their attitudes on indeterminacy and wave-particle duality.

## COURSE BACKGROUND AND TRANSFORMATIONS

Each semester, the University of Colorado (CU) offers two introductory calculus-based modern physics courses; one section is intended for engineering majors (ENG)**,** and the other for physics majors (PHYS). Both versions traditionally cover topics from special relativity and quantum mechanics, with variations from semester to semester according to instructor preferences, and both courses typically enroll ~ 50-100 students. A team from the physics education research (PER) group at CU began developing in 2005 a transformed curriculum for the engineering course [5] that incorporated interactive engagement techniques (in-class concept questions, peer instruction, and computer simulations [6]), as well as revised content intended to emphasize reasoning development, model building, and connections to real-world problems. The progression of quantum physics topics in this course can be broken into three main sections: classical and semi-classical physics; the development of quantum theory; and its application to physical systems.

Informed by our own research into student perspectives, we recently introduced further transformations to this modern physics curriculum, primarily in the middle section of the course. The objectives of these transformations were to: (a) make *realist* expectations explicit; (b) provide *evidence* against those expectations; and (c) attend to student attitudes on interpretive themes across a broad selection of topics. The weekly homework assignments consisted of online submissions and written, long-answer problems; there was a broad mixture of conceptual and calculation problems, both requiring short-essay, multiple-choice, and numerical answers. An online discussion board was created to allow students to anonymously ask questions and provide answers to each other (which afforded us many opportunities to gauge the accessibility of the new material to students). In lieu of a long answer section on the final exam, students were asked to write a 2-3 page (minimum) final essay on a topic from quantum mechanics of their choosing, or to write a personal reflection on their experience of learning about quantum mechanics (an option chosen by ~40% of students). As opposed to a formal term paper, this assignment was meant to give students the opportunity to explore an aspect of quantum mechanics that was of personal interest to them.

Following our treatment of the Bohr model of hydrogen (where the localized existence of electrons is *assumed*), we developed a semi-classical model of atomic magnetic moments, and their classically expected behavior in magnetic fields (Stern-Gerlach experiments).* This topic naturally invokes ample discussion on the counter-intuitive results of repeated spin-projection measurements, the need for probabilistic descriptions in quantum mechanics, and the physical meaning of superposition states. Progressing into distant correlated measurements (*quantum entanglement*, *locality*, *completeness*, *hidden-variables*), we developed working definitions of multiple quantum interpretations (e.g. *Realist, Copenhagen, Matter-Wave* [2,4]), framed in terms of the historic debate between Albert Einstein and Niels Bohr. [8] We then proceeded to engage students in an extended argument (with us, and amongst themselves) against *realist* interpretations of quantum phenomena. This argument was *extended* in two senses: 1) We were able to augment a number of standard topics (e.g., the uncertainty principle, atomic models) with discussions of interpretive themes; and 2) We introduced several entirely new topics (e.g., delayed-choice experiments) that created additional opportunities for students to explore the differences between data, interpretation, and scientific theory. In their end-of-term reflective essays, the topics most frequently cited by students as having influenced their perspectives on quantum physics were the single-quanta experiments with light and matter.

## Single-Quanta Experiments

Single-photon and *delayed-choice* experiments demonstrate the dualistic nature of light, and provide strong evidence for nonlocal interpretations, but are only meaningful to students if the details and results of the experiments are accessible to them. We therefore omitted from our lectures extraneous technical details, while still focusing on the process of designing the experiments and creating an adequate photon source. A guiding principle for this course was to avoid (as much as possible) the expectation for students to accept our assertions as a matter of faith. And so rather than simply describing what the experimentalists had meant to demonstrate, and then informing students of their success, we presented them with actual data from original sources. [9-11] Student discussions on the implications of each of three single-photon experiments were inspired by "clicker questions" interspersed throughout lecture. [FIG. 1]

Double-slit experiments with electrons demonstrate the dualistic nature of matter, in that they are individually detected as localized particles, but collectively form an interference pattern over time. When only one electron is present in the apparatus at a time, we observe the same results, and may *interpret* this as each electron *interfering with itself* as a delocalized wave, and then collapsing to a point in its interaction with the detector. Although this phenomenon may be adequately demonstrated in class using the Quantum Wave Interference PhET simulation, [13] we sought in this course to emphasize connections between theory, interpretation, and experimental evidence, and so augmented these lectures with data from some recently realized double-slit experiments. In 2008, Frabboni, et al. reported their fabrication of a double-slit opening in gold foil on a scale of tens of nanometers (including STM images thereof), and demonstrated electron diffraction with both slits open, as well as its absence with one slit covered. [14] Tonomura, et al. produced a movie [15] that literally shows single-electron detection and the gradual buildup of a fringe pattern. [11] Students from prior modern physics courses had often been skeptical as to whether this experiment (where only a single particle passes through the apparatus at a time) could actually be done in practice – in this way, they were able to observe the phenomenon with their own eyes.

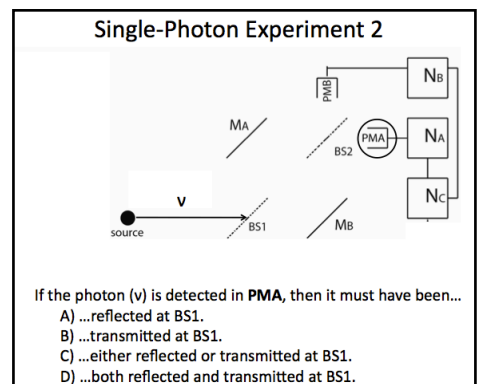

**FIGURE 1.** This sample concept question can serve to generate in-class discussion on the differences between experimental data and a physical interpretation of that data. [BS = Beam Splitter; M = Mirror; PM = Photomultiplier; N = Counter.]

---
* This approach was partly inspired by D. F. Styer. [7]

## COMPARATIVE OUTCOMES

In order to gauge the impact on student thinking, we compare several outcomes from this course incorporating interpretive themes [ENG-INT] with three other recent modern physics offerings at CU. All four courses were large-lecture (N=60–100 for each), utilized interactive engagement in class, and varied in their instructional approaches with respect to interpretation in quantum mechanics. These differences may be best illustrated by how each instructor addressed in class the double-slit experiment with electrons. The *Realist/Statistical* instructor [ENG-R/S] taught that each particle passes through one slit or the other, but that determining which one will disrupt the interference pattern. The *Matter-Wave* instructor [ENG-MW] promoted a wave-packet description where each electron propagates through both slits and then becomes localized upon detection. The *Copenhagen/Agnostic* instructor [PHYS-C/A] touched on interpretive questions, but ultimately emphasized predicting features of the interference pattern (mathematical calculation). ENG-MW is the engineering course most similar to ENG-INT (and to the original transformed curriculum), in that similar lecture materials were used, and interpretive themes were discussed near the end, but in that course without specific reference to atomic systems. PHYS-C/A is a class for physics majors that also used many of the same lecture materials, but with less emphasis on interpretation.

Student interest in quantum mechanics at CU *before* instruction in modern physics is moderately high, at an average between 75-80% favorable. [FIG. 2] However, their *post-instruction* interest typically decreases (to below 70%), with negative responses increasing significantly ($p < 0.001$) – nearly 1/3 of our engineering students would *not* agree that quantum mechanics is an interesting subject after having learned about it in modern physics! This alone seems sufficient reason for introducing further transformations to our typical curriculum. Students from ENG-INT were nearly unanimous (98%) in their reported interest in quantum physics, and not one student responded with a negative opinion. [Relative to the number of students who completed the final exam, the response rate for the ENG-INT post-instruction survey was ~90%.]

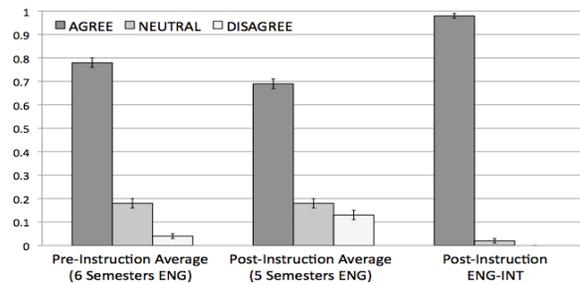

**FIGURE 2.** Average pre- and post-instruction student responses to the statement: *I think quantum mechanics is an interesting subject*. N ~ 50-100 for each semester; error bars represent the standard error on the proportion.

ENG-MW, PHYS-C/A and ENG-INT all offered similar discussions of the Schrödinger model of hydrogen, with a few notable exceptions. Like the first two courses, ENG-INT showed how Schrödinger predicted zero orbital angular momentum for an electron in the ground state, and contrasted this result with the predictions of Bohr and de Broglie. But we continued by arguing how this has implications for the physical interpretation of the wave function – for (as the argument goes) how could conservation of angular momentum allow a localized particle to exist in a state of zero angular momentum in its orbit about the nucleus? More importantly, having already established language and concepts specific to interpretive themes in quantum mechanics, we were able to identify the position of an atomic electron as yet another example of a hidden variable, which we had argued throughout don't exist as a matter of principle. ENG-INT is the only course among these four where a significant majority of students chose at the end of the semester to disagree with the idea of localized atomic electrons. [FIG. 3]

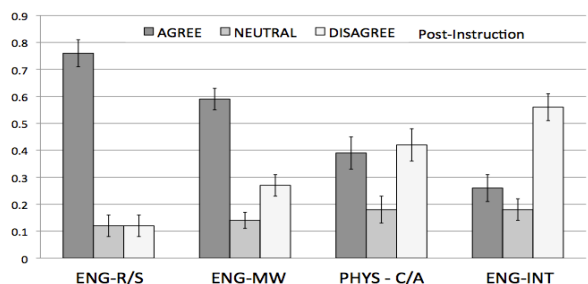

**FIGURE 3.** Post-instruction student responses to the statement: *When not being observed, an electron in an atom still exists at a definite (but unknown) position at each moment in time*. N ~ 50-100 for each course, as denoted in the text; error bars represent the standard error on the proportion.

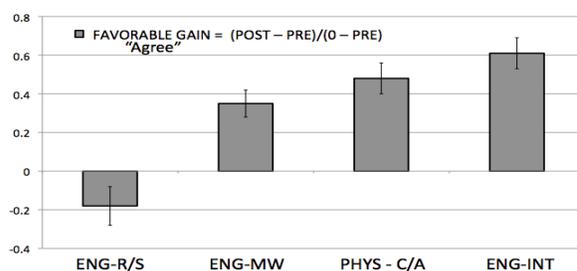

**FIGURE 4.** Favorable gain = (post-pre)/(0-pre) in student agreement with the statement: *The probabilistic nature of quantum mechanics is mostly due to the limitations of our measurement instruments*; where agreement is considered unfavorable. N ~ 50-100 for each course, as denoted in the text; error bars represent the standard error on the proportion.

Even if the physical interpretation of atomic wave functions is not of primary importance for every modern physics instructor, a common learning goal is for students to recognize a difference between the experimental uncertainty of classical mechanics and the fundamental uncertainty of quantum physics. *Realist* expectations might lead pre-instruction students to favor agreement with the statement: *The probabilistic nature of quantum mechanics is mostly due to the limitations of our measurement instruments*. The incoming percentage of students from all three of the engineering courses who agreed with this statement was nearly identical (~45%) while incoming attitudes for the physics majors were significantly more favorable (with only a quarter of them agreeing, and over half disagreeing before instruction).

The differential impact on student responses from these four modern physics courses is most dramatically illustrated by normalizing shifts in student agreement, according to their rate of agreement at the start of the course.[*] [FIG. 4] By this measure, ENG-INT had the greatest positive impact on engineering student attitudes regarding the relationship between fundamental uncertainty in quantum mechanics and classical experimental uncertainty, comparable with the course for physics majors.

## DISCUSSION

A common lament among educators and researchers is that we are losing students in our introductory classical courses by only teaching them physics from the 19th century; similar issues may arise when modern physics instructors limit course content to the state of knowledge in the first half of the last century, when questions of classical and quantum reality were considered to be philosophical in nature. Addressing modern experiments on the foundations of quantum mechanics was overwhelmingly popular among students, and had a demonstrably positive impact on student thinking. We encourage instructors to consider these results when designing their own courses.

## ACKNOWLEDGMENTS

This work was supported in part by NSF CAREER Grant No. 0448176 and the University of Colorado. We thank the students, instructors, and the PER group at CU for making these studies possible.

---

[*] We define *favorable gain* as the negative of this, since a *decrease* in agreement with this statement is considered favorable. This definition is equivalent to the usual *normalized gain* = (post − pre)/(1 − pre), except the target response rate for agreement is zero instead of 100%.